\documentclass[%
 reprint,
 amsmath,amssymb,
 aps,
prl,
]{revtex4-1}

\usepackage{mathtools}
\usepackage{graphicx}
\usepackage{dcolumn}
\usepackage{bm}

\usepackage{color}

\allowdisplaybreaks

\begin{document}



\preprint{APS/123-QED}

\title{Generative framework for dimensionality reduction of large 
scale network of non-linear dynamical systems driven by external input}


\author{Shrey Dutta}
\email{0shreydutta0@gmail.com}
\author{Dipanjan Roy}%
\email{dipanjan.nbrc@gov.in}
\author{Arpan Banerjee}
\email{arpan@nbrc.ac.in}

\affiliation{%
 National Brain Research Centre, Manesar,Gurgaon, Haryana-122052, India\\
}%


\begin{abstract}
Several studies have proposed constraints under which 
a low dimensional representation can be derived from large-scale 
real-world networks exhibiting complex nonlinear dynamics. Typically, 
these representations are formulated under certain assumptions, such as 
when solutions converge to attractor states using linear stability analysis 
or using projections of large-scale dynamical data into a set of lower dimensional 
modes that are selected heuristically. Here, we propose a generative framework for 
selection of lower dimensional modes onto which the entire network dynamics can be 
projected based on the symmetry of the input distribution for a large-scale network 
driven by external inputs, thus relaxing the heuristic selection of modes made in the 
earlier reduction approaches. The proposed mode reduction technique is tractable 
analytically and applied to different kinds of real-world large-scale network scenarios 
with nodes comprising of a) Van der Pol oscillators b) Hindmarsh-Rose neurons. These 
two demonstrations elucidate how order parameter is conserved at original and reduced 
descriptions thus validating our proposition.
\end{abstract}

\pacs{Valid PACS appear here}
\maketitle


Large-scale dynamical systems are useful tools to explain a wide variety of complex phenomena 
in nature e.g. financial markets \cite{DavidFM}, jamming transitions \cite{charboJamming}, 
human mobility dynamics \cite{songHuman}, weather patterns \cite{AntoulasBeattieGugercin2010} 
and brain dynamics \cite{DecoJirsaRobinsonEtAl2008}. While increase in scale or dimensions 
may increase the predictive power of the model system, nonetheless a reduction to simpler 
descriptions at lower dimensions is critical for having relevance to empirical observations 
and analytical tractability of underlying mechanisms governing empirical observations. One 
robust approach of reducing dimensions is defining modes on which the original system can 
be projected \cite{AntoulasBeattieGugercin2010,SJ3D2008}. The selection of a mode is often 
heuristically motivated, and the mode can also be an order parmeter from the perspective 
of the paradigmatic framework of Synergetics \cite{synergetics}. In Neuroscience, reduction of 
dynamical systems with respect to modes constructed from distribution of external input has been 
performed earlier on small-scale network of linearly coupled excitable systems \cite{SJ3D2008}. 
Since this reduction retain important network dynamics, large-scale networks were conceptualized 
by coupling these reduced systems with long-range coupling \cite{becker2014, paula2015}, the later 
being heuristically argued from symmetry properties. In present work, we perform reduction 
on a large-scale network where connection among nodes involve global and local coupling mimicking 
a real-world system. Subsequently, long-range coupling term between modes in the reduced system 
is derived analytically as part of the reduction process. Global coherence is an order parameter 
that can be computed both at the level of original dynamical system as well as from the mode 
dynamics in the reduced system. Conservation of global coherence at both levels is used to validate 
the generality of our approach in two distinct networks. First, we simulate a large-scale network 
where each node is a Van der Pol oscilator having 2-dimensional dynamics and coupled using local 
and global parameters. Second each node is a Hindmarsh-Rose neuron, a 3-dimensional dynamical 
system which can exhibit different time scales of oscillations resulting in bursting along with 
tonic spiking behavior.

\begin{figure}
	\centering
	\includegraphics[scale =0.2]{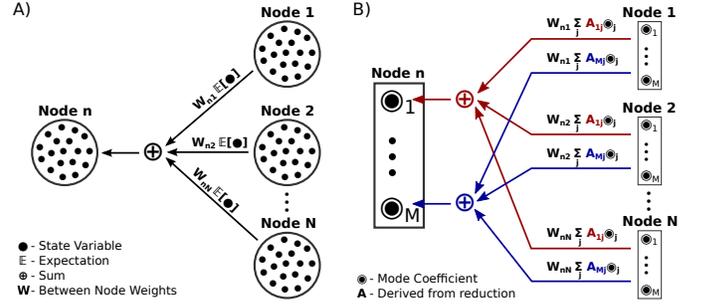}
	\caption{\label{fig1}A) Large-scale network architecture of complete
		system where input to a node is the weighted sum of expected value of
		state-variables B) Large-scale network architecture of 
		proposed reduced system where input to a target-mode
		is weighted sum of target-mode specific expected mode activity
		from other nodes.   }
\end{figure}

Each $i^{th}$ excitable system in node $n$ is represented by a vector of state variables (state vector) $\bm{x}^{(n)}_i$. Input to $i^{th}$ excitable system from interaction with other excitable systems within $n^{th}$ node is given by a vector function $\bm{g}\left(\{\bm{x}^{(n)}_{j}\}, K, i\right)$ where $K$ is coupling constant and $\{\bm{x}^{(n)}_{j}\}$ is a set of $J$ state-vectors ($j = 1,...,J$) in $n^{th}$ node. Input to every excitable system in $n^{th}$ node from other nodes is given by another vector function $\bm{h}\left(\{\bm{x}_j^{(m)}\}, W, n\right)$ where elements of matrix $W$ ($W\in\Re^{J \times N}$) are the weights of connections between nodes and $\{\bm{x}_j^{(m)}\}$ is a set of state-vectors in $N$ nodes ($m=1,...,N$) (Figure \ref{fig1}A). Vector function 
$\bm{f}(\bm{x}_i^{(n)})$ contributes to the local dynamics of $i^{th}$ excitable system in $n^{th}$ node. Then, the dynamics of the entire system is described by the following set of equations 
\begin{eqnarray}
\label{eq1}
\tau_n \dot{\bm{x}}^{(n)}_i = && \bm{f}\left(\bm{x}_i^{(n)}\right) + \bm{g}\left(\{\bm{x}^{(n)}_{j}\}, K, i \right) + \nonumber\\ 
&& \bm{h}\left(\{\bm{x}_j^{(m)}\}, W, n\right) + \bm{\phi}\left(\bm{I}_i^{(n)}\right)
\end{eqnarray}
where 
$\dot{\bm{x}}$ is time derivative of state vectors, $\bm{\phi}$ is a function of the  external input ($\bm{I}$) and $\tau_n$ is the time-constant of $n^{th}$ node which is a differentiating factor between nodes. However, within $n^{th}$ node, external current to $i^{th}$ excitable system ($\bm{I}^{(n)}_i$)  differentiates it from the rest. For a large scale network comprising of individual excitable nodes, \eqref{eq1} can be written as 
\begin{eqnarray}
\label{eq2}
\tau_n \dot{\bm{x}}^{(n)}(t, \bm{I}) = && \bm{f}\left(\bm{x}^{(n)}\right) + \bm{g}\left(\bm{x}^{(n)}, K \right) + \nonumber\\ 
&& \bm{h}\left(\{\bm{x}^{(m)}\}, W, n,i\right) + \bm{\phi}\left(\bm{I}\right)
\end{eqnarray}
where $I$ is a continuous variable having normal distribution $\mathbb{N}(\mu,\sigma)$.

Now, we can represent $\bm{x}(t, I)$ as a superposition of $M$ bi-orthogonal modes $\{v_i\}$

\begin{eqnarray}
\label{eq3}
\bm{x}(t, \bm{I}) = \sum_{i=1}^M \bm{\xi}_i(t)v_i(\bm{I}) + R(t,\bm{I})
\end{eqnarray}
where $R(t,\bm{I})$ is the residual and $M\lll J$ (Figure \ref{fig1}B). The nature of reduction is such that dynamical system given in \eqref{eq1} is reduced to solving for the mode coefficients $\bm{\xi}_i$ as described in the following set of equations 
\begin{eqnarray}
\label{eq4}
\tau_n \dot{\bm{\xi}}^{(n)}_i = && \bm{F}\left(\bm{\xi}_i^{(n)}\right) + \bm{G}\left(\{\bm{\xi}^{(n)}_{j}\}, K, i \right) + \nonumber\\ 
&& \bm{H}\left(\{\bm{\xi}_j^{(m)}\}, W, n, i\right) + \bm{II}_i^{(n)}
\end{eqnarray} where
\begin{eqnarray}
\bm{\xi}_i = && \int_I \bm{x}(t,I)v^+_i dI \nonumber \\
\bm{F}\left(\bm{\xi}_i\right) = && \int_I \bm{f}\left(\bm{x}\right) v_i^+ dI \nonumber \\
\bm{G}\left(\{\bm{\xi}_{j}\}, K, i \right) = &&\int_I  \bm{g}\left(\bm{x}, K \right) v_i^+ dI \nonumber \\
\bm{H}\left(\{\bm{\xi}^{(m)}_{j}\}, W, n,i \right) = &&\int_I  \bm{h}\left(\{\bm{x}^{(m)}\}, W,n,i\right) v_i^+ dI \nonumber \\
\bm{II}_i^{(n)} = && \int_I \bm{I} v_i^+ dI\nonumber
\end{eqnarray}
and $\{v_i^+\}$ are the adjoint basis for the biorthogonal 
modes $\{v_i\}$.
\\

\noindent
\textbf{Large-scale network of Van der Pol oscillators:}

A Van der Pol oscillator \cite{VanDerPol1920} has two state variables $x$ and $y$ which follows the following equations
\begin{eqnarray}
\dot{x} = && y \nonumber \\
\dot{y} = && -a(x^{2}-1)y - x
\end{eqnarray}
A large-scale network where individual node is essentially a Van der Pol oscillator can be cast into equation \eqref{eq1} with the following relations
\begin{eqnarray}
\bm{f}\left(\bm{x}_i\right) = &&\left[ 
	\begin{array}{c} 
		y_i \\
                -a({x_i}^2-1)y_i - x_i
        \end{array} \right] \nonumber \\
\bm{g}\left(\{\bm{x}_{j}\}, K, i \right)  = &&\left[ 
	\begin{array}{c} 
		K (\mathbb{E}[\{x_j\}] - x_i) \\
                0
        \end{array} \right] \nonumber \\
\bm{h}\left(\{\bm{x}_j^{(m)}\}, W, n\right)= &&\left[ 
	\begin{array}{c} 
		\sum_{m =1}^N W_{nm} \mathbb{E}[\{x^{(m)}_j\}] \\
                0
        \end{array} \right] \nonumber \\ 
\bm{I} = && \left [
	\begin{array}{c}
	 I \\
	 0
	\end{array} \right] \nonumber
\end{eqnarray} 
where $a$ is a constant and $\mathbb{E}[\{x_j\}]$ is the expected value of the state-variables $\{x_j\}$. For the reduced system  described in \eqref{eq4} with $\xi_{i} = [\alpha_{i},\beta_{i}]^{T}$, we derive

\begin{eqnarray}
\bm{F}\left(\bm{\alpha}_i\right) = &&\left[ 
	\begin{array}{c} 
		\beta_i \\
                -a_i{\alpha_i}^2\beta_i + a \beta_i - \alpha_i
        \end{array} \right] \nonumber \\
\bm{G}\left(\{\bm{\alpha}_{j}\}, K, i \right) = &&\left[ 
	\begin{array}{c} 
		K \left(\sum_{j=1}^{M} A_{ij}\alpha_j - \alpha_i \right)\\
                0
        \end{array} \right] \nonumber \\
\bm{H}\left(\{\bm{\alpha}^{(m)}_{j}\}, W, n,i \right) = &&\left[ 
	\begin{array}{c} 
		\sum_{m =1}^N W_{nm} \sum_{j=1}^{M} A_{ij}\alpha^{(m)}_j\\
                0
        \end{array} \right] \nonumber \\
\bm{II} = && \left [
	\begin{array}{c}
	 II \\
	 0
	\end{array} \right] \nonumber
\end{eqnarray} where the constants $a_i$ and $A_{ij}$ are computed by applying bi-orthogonal assumption of modes as stated in Appendix A.  \\

\noindent
\textbf{Large-scale network of Hindmarsh-Rose neurons:}

In Neuroscience, Hindmarsh-Rose (HMR) neuron is a three dimensional model of single neuron firing dynamics having three state variables \cite{HMR84} 
\begin{eqnarray}
\dot{x} = && y - ax^{3} + bx^{2}+I-z \nonumber \\
\dot{y} = && c-dx^{2}-y \nonumber \\
\dot{z} = && r(s(x-x0)-z) 
\end{eqnarray} where $a$, $b$, $c$, $d$, $r$, $s$ and $x0$ are the constants. A network of excitatory and inhibitory HMR neurons have been used to describe a small-scale network of neurons \cite{SJ3D2008}. Thus, a node in the brain (Fig \ref{fig1}a) can be expressed as a six dimensional state space with 3 excitatory and 3 inhibitory variables represented as $\bm{x} = \left[x_1, y_1, z_1, x_2, y_2, z_2 \right]^T$. Extending this architecture, for the large-scale system in \eqref{eq1}, we obtain

\begin{eqnarray}
\bm{f}\left(\bm{x}_i\right) = &&\left[ 
	\begin{array}{c} 
		{y_1}_i - a {{x_1}_i}^3 + b {{x_1}_i}^2 - {z_1}_i \\
                c - d {{x_1}_i}^2 - {y_1}_i\\
                r(s({x_1}_i - x0) - {z_1}_i)\\
         	{y_2}_i - a {{x_2}_i}^3 + b {{x_2}_i}^2 - {z_2}_i \\
                c - d {{x_2}_i}^2 - {y_2}_i\\
                r(s({x_2}_i - x0) - {z_2}_i)                       
	\end{array} \right] \nonumber \\ 
\bm{g}\left(\{\bm{x}_{j}\}, K, i \right) =  &&\left[ 
	\begin{array}{c} 
		\left( \begin{array}{c}
		  K_{11} (\mathbb{E}[\{{x_1}_j\}] - {x_1}_i) - \\
		 K_{12} (\mathbb{E}[\{{x_2}_j\}] - {x_1}_i)
		\end{array}\right) \\
                0 \\
		0 \\
		K_{21} (\mathbb{E}[\{{x_1}_j\}] - {x_2}_i) \\
		0 \\ 
		0 		
        \end{array} \right] \nonumber \\
\bm{h}\left(\{\bm{x}_j^{(m)}\}, W, n\right)= &&\left[ 
	\begin{array}{c} 
		\sum_{m =1}^N W_{nm} \mathbb{E}[\{{x_1}^{(m)}_j\}] \\
                0\\
		0\\
		0\\
		0\\
		0
        \end{array} \right] \nonumber \\ 
\bm{I} = && \left [
	\begin{array}{c}
	 I1 \\
	 0 \\
	 0 \\
	 I2\\
	 0\\
    	 0
	\end{array} \right] \nonumber
\end{eqnarray}

where $K_{11}$, $K_{12}$ and $K_{21}$ are  coupling constants between excitatory and inhibitory nodes. For the reduced system in \eqref{eq4}, $\xi_{i} = [\alpha_{1},\beta_{1},\gamma_{1},\alpha_{2},\beta_{2},\gamma_{2}]^{T}$, we derive

\begin{eqnarray}
\bm{F}\left(\bm{\xi}_i\right) = &&\left[ 
	\begin{array}{c} 
		{\beta_1}_i - {a_1}_i {{\alpha_1}_i}^3 + {b_1}_i {{\alpha_1}_i}^2 - {\gamma_1}_i \\
                {c_1}_i - {d_1}_i {{\alpha_1}_i}^2 - {\beta_1}_i\\
                rs{\alpha_1}_i - r{\gamma_1}_i - {p_1}_i\\
         	{\beta_2}_i - {a_2}_i {{\alpha_2}_i}^3 + {b_2}_i {{\alpha_2}_i}^2 - {\gamma_2}_i \\
                {c_2}_i - {d_2}_i {{\alpha_2}_i}^2 - {\beta_2}_i\\
                rs{\alpha_2}_i - {\gamma_2}_i - {p_2}_i                       
        \end{array} \right] \nonumber \\
\bm{G}\left(\{\bm{\xi}_{j}\}, K, i \right)  = &&\left[ 
	\begin{array}{c} 
		\left(\begin{array}{c}
		  K_{11} \left(\sum_{j=1}^{M} A_{ij}{\alpha_1}_j - {\alpha_1}_i \right) - \\
		  K_{12} \left(\sum_{j=1}^{M} B_{ij}{\alpha_2}_j - {\alpha_1}_i \right)
		\end{array} \right) \\
                0 \\
		0 \\
		K_{21} \left(\sum_{j=1}^{M} C_{ij}{\alpha_1}_j - {\alpha_2}_i \right) \\
		0 \\ 
		0 \\
        \end{array} \right] \nonumber \\
\bm{H}\left(\{\bm{\xi}^{(m)}_{j}\}, W, n,i \right) = &&\left[ 
	\begin{array}{c} 
		\sum_{m =1}^N W_{nm} \sum_{j=1}^{M} A_{ij}{\alpha_1}^{(m)}_j\\
                0 \\
		0 \\
		0 \\
		0 \\
		0
        \end{array} \right] \nonumber \\ 
\bm{II} = && \left [
	\begin{array}{c}
	 II1 \\
	 0 \\
	 0 \\
	 II2\\
	 0\\
    	 0
	\end{array} \right] \nonumber
\end{eqnarray}

where the constants ${a_1}_i$, ${a_2}_i$, ${b_1}_i$, ${b_2}_i$, ${c_1}_i$, ${c_2}_i$, 
${d_1}_i$, ${d_2}_i$, ${p_1}_i$, ${p_2}_i$, $A_{ij}$, $B_{ij}$ and $C_{ij}$ are 
defined in Appendix.


\begin{figure}
\centering
\includegraphics[scale=0.45]{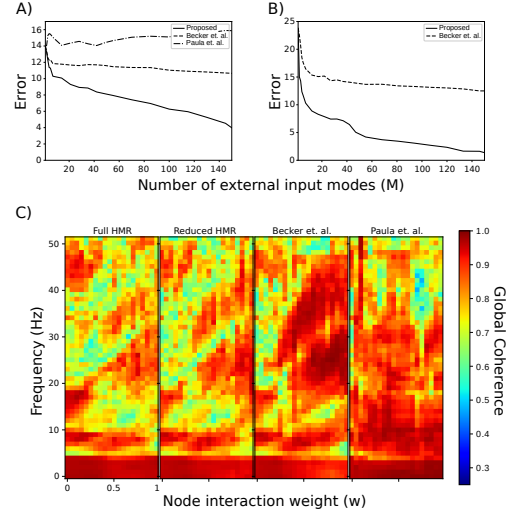}
\caption{\label{fig2}A) Error computed 
	for Hindmarsh-Rose model for $a = 1$,$b=3$, $c=1$, $d=5$, $r=0.006$, $s=4$, $x0=-1.6$, $K11=0.5$, $K21=0.5$, $Inhibition-excitation ratio (IER) \in (0,2.5)$, $\mu \in (0,2.5)$, $\sigma=0.4$, $w\in(0,1)$, $M \in (1,150)$, $\tau_1 = 0.05ms$, $\tau_2 = 1ms$ and $\tau_3 = 2.5ms$ B) Error computed for VanderPol model for $a = 0.1$, $K \in (0,0.5)$, $\mu \in (0,0.5)$, $w \in (0,1)$, $\tau_1 = \tau_2 = \tau_3 = 0.05 ms$, $\sigma = 0.4$ and $M \in (1,150)$ C) Global coherence spectra is plotted for varying node interaction levels in the large-scale network for Hindmarsh-Rose model with $\mu = 1.75$ and $IER = 2.0$ for full and reduced cases (30 modes).}
\end{figure}

We simulated a network of three nodes with each functional unit in a node is governed by 1) Van der Pol (VDP) oscillator or 2) HMR neurons. We select the node connection matrix of the given form
\begin{eqnarray}
W = \begin{bmatrix}
		 0 &1 &1 \\
		-1 &0 &1\\
		-1 &-1 &0
	\end{bmatrix} \times w \nonumber
\end{eqnarray}
where $w$ is a scalar in the range $0$ to $1$ representing  negligible node interaction to strong 
node interaction. We also considered different values of $K$ for VDP (and Inhibition-excitation 
ratio, $IER = \frac{K_{12}}{K_{11}}$ for HMR) and for different values of mean current $\mu$. For validating reduced system with $M$ modes, we compare the  Global Coherence (GC) evaluated using the complete system ($C_p$) with GC evaluated using the reduced system ($R_p$) using the following equation.
\begin{eqnarray}
\label{eq5}
error(M) = \sqrt{\sum_{p=<f,w,K,\mu>}^T \left( 
		\begin{array}{c}
		C_{p} - R_{p} 
		\end{array}
	   \right)^2
	   }
\end{eqnarray}
where $f$ is frequency and  $T$ is total number of pairs $<f,w,K,\mu>$. GC is computed between mean activities of each node. For the reduced system, the mean activity of each node is the mean of individual excitable systems' activities estimated using 
\eqref{eq3} without the residual.

\begin{figure}
	\centering
	\includegraphics[scale =0.25]{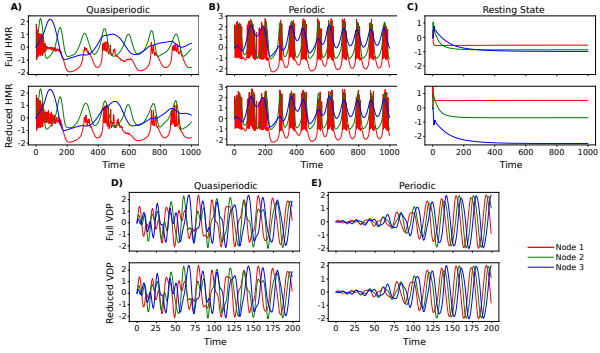}
	\caption{\label{fig3}Preservation of mean-field activity of each node in the reduced model 
	with 30 modes for Hindmarsh-Rose model A) $w =0.5$, $\mu = 2.5$, $IER = 0.8$ and $K11 = 0.5$ B) $w=0.5$, $\mu = 2.5$, $IER = 0.8$ and $K11 = 30$ and C) $w = 0.5$, $\mu = 2.5$, $IER = 2.5$ and $K11 = 30$ and for Vander-Pol model D) $w = 0.5$, $K = 0.1$ and $\mu = 1$  and E) $w = 0.5$, $K = 0.1$ and $\mu = 0.1$.}
\end{figure}

As $M$ increases, error in proposed reduction process decreases more rapidly as compared to 
previous approaches (Figure \ref{fig2}A \& B) for both large-scale networks using Van der-Pol and 
Hindmarsh-Rose models as nodes, thus, validating our reduction approach. For Van der-Pol model, 
reduction proposed in \cite{paula2015} generated numerical instability during simulation. Global 
coherence calculated from proposed reduction for an exemplar parameter space matches closely 
with the 
original system unlike heuristic approaches (Figure \ref{fig2}C). We further 
validate our framework by showing the reproduction of time-series of mean-field activity of 
each HMR and VDP node (Figure \ref{fig3}) for several different parameters spaces.

%
%
%
%
%


In \cite{SJ3D2008}, local-interaction between state-variables was facilitated via the mean field 
activity of the node which by itself is a small-scale network. In the reduced model this 
interaction was preserved as the input to state variable of mode $i$ was characterized 
via mode-specific output of the node ($\sum_{j=1}^M A_{ij}\alpha_j$) governed by matrix $A$ which 
is obtained as a part of reduction. However, in the case of the large-scale network of these reduced 
nodes the input to state-variable of mode $i$ of node $m$ was either the activity of state-variable 
of mode $i$ from other nodes \cite{becker2014} or it was the sum of activities of state-variables 
of all modes from other nodes \cite{paula2015}. In this paper, the input to mode $i$ of node $m$ is 
derived by projecting the long-range interaction term of complete large-scale network 
to modes of the external input. Thus, our proposed scheme preserves the original long range 
interaction (Figure \ref{fig2} and Figure \ref{fig3}).

In summary, we propose a generalized scheme for reduction of the dynamics of a large-scale network
into lower dimensional mode description based on properties of the external input. Obviously, any 
such reduction lowers the computational complexity. However an important point to note is a 
model's benefit is not necessarily limited to mimicking the complex dynamics of real-world 
system. For example, a detailed model of the cortical layer will be highly informative 
\cite{Markram2015}, but not necessarily insightful for explaining the cortical interactions 
during a specific behavioral task. How do we develop task-specific insights from a high 
dimensional data with minimal assumptions will be critical question for future studies and 
how they could profit from our proposed formalism. Our approach is best-placed to application 
where task inevitably involves reduction from high-dimensional state space to functionally well 
defined lower dimensional modes. Such problems are also not necessarily exclusively brain 
specific, but also pertinent for climate dynamics, traffic problems, or anywhere dynamical 
systems are driven by external time varying and distributed input and the approach presented 
in this article may be insightful for a wide range of scientific disciplines.

This research was funded by NBRC core and the grants Ramalingaswami fellowship 
(BT/RLF/Re-entry/31/2011) and Innovative Young Bio-technologist Award (IYBA) 
(BT/07/IYBA/2013) from the Department of Biotechnology (DBT), Ministry of Science \& Technology, 
Government of India to AB. DR is supported by the Ramalingaswami fellowship 
(BT/RLF/Re-entry/07/2014) and DST extramural grant (SR/CSRI/21/2016). 
We also thank Prof. Michael Breakspear for helpful comments and discussions.
\\
\appendix

\noindent
\textbf{Appendix A: Reduction coefficients for Van der Pol oscillator network}

\begin{eqnarray}
A_{ij} &= \int_{I^\prime} g(I^\prime) v_j(I^\prime) dI^\prime \int_I v_i^+ dI \nonumber \\ 
a_i &= a  \int_I v_i^3(I) v_i^+(I) dI \nonumber  \\
II_i &=  \int_I I v_i^+(I) dI \nonumber \\
\end{eqnarray}
$g(I)$ is the pdf of external input.
\\

\noindent
\textbf{Appendix B: Reduction coefficients for Hindmarsh-Rose neuronal network}
\begin{subequations}
\begin{eqnarray}
A_{ij} &= \int_{I^\prime} g1(I^\prime) v1_j(I^\prime) dI^\prime \int_I v1_i^+(I) dI \nonumber \\ 
B_{ij} &= \int_{I^\prime} g2(I^\prime) v2_j(I^\prime) dI^\prime \int_I v1_i^+(I) dI \nonumber \\
C_{ij} &= \int_{I^\prime} g1(I^\prime) v1_j(I^\prime) dI^\prime \int_I v2_i^+(I) dI \nonumber \\
a1_i &= a  \int_I v1_i^3(I) v1_i^+(I) dI \nonumber \\
a2_i &= a  \int_I v2_i^3(I) v2_i^+(I) dI \nonumber \\
b1_i &= b  \int_I v1_i^2(I) v1_i^+(I) dI \nonumber \\
b2_i &= b  \int_I v2_i^2(I) v2_i^+(I) dI \nonumber \\
c1_i &= c  \int_I v1_i^+(I) dI \nonumber\\
c2_i &= c  \int_I v2_i^+(I) dI \nonumber\\
d1_i &= d  \int_I v1_i^2(I) v1_i^+(I) dI \nonumber \\
d2_i &= d  \int_I v2_i^2(I) v2_i^+(I) dI \nonumber\\
p1_i &= r s x_0 \int_I v1_i^+(I) dI \nonumber \\
p2_i &= r s x_0 \int_I v2_i^+(I) dI \nonumber\\
II1_i &=  \int_I I v1_i^+(I) dI \nonumber\\
II2_i &=  \int_I I v2_i^+(I) dI \nonumber
\end{eqnarray}
\end{subequations}

$g1(I)$ and $g2(I)$ are the pdfs of external input to excitatory and inhibitory sub-populations respectively.

\bibliography{reduced_network}

\end{document}